\documentclass[preprint,12pt]{elsarticle}
\usepackage{amsmath,amssymb,graphicx, bm,psfrag,amsfonts,theorem}
\newtheorem{theorem}{Theorem}
\newtheorem{proposition}[theorem]{Proposition}

\psfrag{lambda}[t]{$\lambda$}
\psfrag{rho1}[l][l][0.9]{$\rho^{\mathrm{MP}}$}
\psfrag{rho2}[l][l][0.9]{$\rho^{\mathrm{WBE}}$}
\psfrag{capacity}[l][l][0.9][0]{$C$ [bit/user/Hz]}
\psfrag{snr}[t]{SNR [dB]}
\begin{document}
\begin{frontmatter}



\title{Optimization of sequences in CDMA systems: a statistical-mechanics approach
}


\author[label1]{Koichiro Kitagawa}
\ead{kitagawa@sys.i.kyoto-u.ac.jp}
\author[label1]{Toshiyuki Tanaka}
\ead{tt@i.kyoto-u.ac.jp}
\address[label1]{Graduate School of Informatics, Kyoto University, Japan.\fnref{cor1}}
\fntext[cor1]{Graduate School of Informatics, Kyoto University, 36-1 
Yoshida Hon-machi, Sakyo-ku, Kyoto-shi, Kyoto 606-8501, Japan.}

\begin{abstract}
Statistical mechanics approach is useful not only in analyzing macroscopic system performance
of wireless communication systems, 
but also in discussing design problems of wireless communication systems.
In this paper, we discuss a design problem of spreading sequences 
in code-division multiple-access (CDMA) systems, 
as an example demonstrating the usefulness of statistical mechanics approach. 
We analyze, via replica method, the average mutual information between inputs and outputs 
of a randomly-spread CDMA channel, and discuss the optimization problem with 
the average mutual information as a measure of optimization.  
It has been shown that the average mutual information is maximized by 
orthogonally-invariant random Welch bound equality (WBE) spreading sequences.
\end{abstract}


\end{frontmatter}


\section{Introduction}
In recent years, advances in information and communication technologies 
have been demanding high data-rate wireless communications.  
In order to realize high data-rate wireless communications, bandwidth of systems should be as wide as possible, 
which means that such systems should have a large degree of freedom.
Those systems are also required to be able to operate efficiently 
even in bad and uncertain environments.
For example, in urban areas, there are many obstacles, such as buildings, cars, and people, 
which interact with wireless communication systems as reflecting/scattering bodies, 
thereby making communication environment very complex.
In analyzing wireless communication systems, therefore, 
one has to regard them 
as systems with very high dimensionality and randomness.  
This is why statistical mechanics approach is expected to be 
useful in studying wireless communication systems.

In this paper, we consider a problem arising from 
considerations of multiple-access channels. 
Typically, a wireless communication system has to accommodate 
multiple users simultaneously.  
In such a system, signals coming from different users 
interfere with each other.  
How to mitigate such multiple-access interference (MAI) 
is one of the most important problems in wireless communications.  
Code-division multiple-access (CDMA)~\cite{Viterbi,cdma} 
provides an effective scheme to mitigate MAI, 
and is widely used in various commercial systems.  
In CDMA, an information symbol of a user is modulated with 
a spreading sequence assigned to the user.
Receiver has to estimate information symbols based on received sequences 
by utilizing knowledge of spreading sequences of the users.  

A conventional choice to mitigate MAI is to use pseudorandom sequences as the spreading sequences.  
Although analysis of such randomly-spread CDMA systems was thought to be difficult, 
it has turned out that {\it replica method}, which is an analytical tool developed in the research field 
of statistical physics of disordered systems (spin glasses), is very useful for the analysis~\cite{tt,guo}.  
More precisely, these studies have revealed that 
replica method allows us to evaluate ``macroscopic'' performance 
of CDMA systems with independent and identically-distributed (i.i.d.) 
random spreading sequences in the large-system limit, 
such as mutual information between inputs and outputs, bit error rate, and so on.

Since theoretical performance of CDMA systems is affected by 
choices of spreading sequences, 
design of spreading sequences is an important problem in CDMA.  
There have been several researches in which 
the problem of designing spreading sequences 
is formulated in terms of an optimization problem. 
For example, Rupf and Massey~\cite{Rupf} discussed optimization 
of spreading sequences with the channel capacity of CDMA systems 
as a measure of optimization.
They showed that so-called Welch bound equality (WBE) spreading sequences, which minimize 
the total squared correlation (TSC) of spreading sequences, achieve the channel capacity.

We restrict ourselves to considering the optimization problem 
of spreading sequences of CDMA systems 
with the mutual information between inputs and outputs 
as the measure of optimization.  
There have been several researches in which 
such optimization problems are discussed~\cite{Rupf,VA,LUE}.  
They have assumed inputs of the system to be Gaussian distributed 
and discussed theoretical upper bound (i.e., channel capacity) 
of mutual information between inputs and outputs.
If one wishes to consider realistic wireless communication systems, 
however, it is important to study the optimization problem 
under the assumption of non-Gaussian inputs.  
The objective of this paper is therefore 
to discuss the optimization problem of spreading sequences 
of CDMA systems when one allows non-Gaussian inputs.  
We would like to emphasize that, 
unlike previous statistical-mechanics studies 
of CDMA systems~\cite{tt, guo, tuk} 
whose objectives are basically to analyze macroscopic system performance, 
we show in this paper that the statistical-mechanics approach 
is also useful in dealing with design problems 
in wireless communication, with the optimization problem 
of spreading sequences of CDMA systems as a demonstrative example.  
A digest version of this paper has been presented 
as a conference paper~\cite{KT}.
\section{Problem}
We consider the following real-valued $K$-user CDMA channel model, 
\begin{equation}
y_{\mu}= \frac{1}{\sqrt{L}} \sum_{k=1}^K s_{\mu k} x_k + \sigma n_\mu, 
\quad \mu=1,\,\ldots,\,L,
\label{eq:model1}
\end{equation} 
where $x_k$ is an information symbol of user $k$.  
We assume that $\{x_k;\,k=1,\,\cdots,\,K\}$ are 
i.i.d.\ random variables, 
and let $p(\cdot)$ be the prior probability of $x_k$, 
whose mean and variance are assumed to be zero and one, respectively.  
$\{ s_{\mu k};\,\mu=1,\,\cdots,\,L\}$ is the spreading sequence of user $k$
 in the symbol interval of interest, 
and $L$ denotes the spreading factor of the CDMA channel model.  
We assume that the power of the spreading sequences is 
normalized to one, so that 
$\sum_{\mu=1}^L (s_{\mu k}/\sqrt{L})^2 =1$ holds for $k=1,\cdots, K$.  
We assume additive white Gaussian noise (AWGN):
$n_{\mu} \sim \mathcal{N}(0,1)$ so that 
$\sigma^2$ is the variance of AWGN.  
Let us introduce the following notations: 
$\bm{y}\equiv [y_1,\cdots,y_L]^T$, 
$\bm{n}\equiv [n_1,\,\cdots,\,n_{L}]^T$,
$\bm{x}\equiv[x_1,\,\cdots,\,x_K]^T$, and 
$S=(S_{\mu k})$, $S_{\mu k}\equiv (1/\sqrt{L})s_{\mu k};\,k=1,\,\cdots,\,K;\;
\mu=1,\,\cdots,\,L$.   
The system model~\eqref{eq:model1} is then rewritten as
\begin{equation}
\bm{y}=S \bm{x} + \sigma \bm{n}. \label{eq:model2}
\end{equation}

One can consider a maximization problem of per user mutual information 
between $\bm{x}$ and $\bm{y}$ with respect to the spreading sequences $S$, 
with the channel input $\bm{x}$ drawn from the probability distribution 
$p(\bm{x})=\prod_{k}p(x_k)$, 
\begin{equation}
C_{\mathrm{user}} = \frac{1}{K}I(\bm{x};\bm{y})|_S, \label{eq:cap_fin}
\end{equation}
where the notation $I(\bm{x};\bm{y})|_S$ denotes 
the mutual information between $\bm{x}$ and $\bm{y}$ 
when $S$ is specified.
When $K\le L$, 
the mutual information $C_{\mathrm{user}}$ is maximized by assigning to all users 
orthogonal $L$-dimensional vectors as spreading sequences, 
regardless of the input distribution $p(\cdot)$.
When $K > L$, on the other hand, spreading sequences maximizing the mutual information $C_{\mathrm{user}}$ are not trivial.
When $x_k$, $k=1,\,\cdots,\,K$, are i.i.d.\ standard Gaussian random variables, 
it is known that the WBE spreading sequences
maximize the mutual information $C_{\mathrm{user}}$~\cite{Rupf}. 
WBE spreading sequences are characterized as~\cite{Massey, Viswanath}
\begin{equation}
SS^T=\beta I_{L \times L},\quad \beta \equiv \frac{K}{L}>1, \label{eq:WBE}
\end{equation}
where $I_{L \times L}$ is an $L$-dimensional identity matrix.

When one assumes Gaussian inputs, spreading sequences maximizing $C_{\mathrm{user}}$ 
have been identified in more general system models than~\eqref{eq:model2}.
For example, in a system model 
where one allows the power of inputs to be different, 
Viswanath and Anantharam~\cite{VA} showed that the mutual information is maximized by 
assigning orthogonal spreading sequences to relatively high-power users 
and so-called generalized WBE spreading sequences to 
the remaining users, 
where the users are classified according to a certain criterion.  
Also, in a system model where the inputs may arrive asynchronously, 
Luo {\it et al.}~\cite{LUE} showed that the mutual information is maximized by 
spreading sequences which can be regarded as an extension of the ones 
which Viswanath and Anantharam proposed.

On the other hand, when $x_k$'s are drawn from a non-Gaussian distribution, 
to the authors' knowledge, spreading sequences maximizing 
the mutual information~\eqref{eq:cap_fin} have not been known.
We analyze, via statistical mechanics, 
spreading sequences maximizing the mutual information 
of the system with non-Gaussian inputs.
Since the case with non-Gaussian inputs is difficult 
to analyze analytically, 
we resort to making several assumptions.  
First, we evaluate mutual information in the large-system limit, in which the number of users $K$ 
and the spreading factor $L$ are both sent to infinity while maintaining their ratio $\beta=K/L$ constant.
Second, we assume random spreading.  
More specifically, we assume that the sample correlation matrix $R=S^TS$ of 
random spreading sequences $S$ is asymptotically orthogonally invariant, 
that is, the probability law of $R$ and that of an orthogonal transform $U^TRU$ 
are the same for any orthogonal matrix $U$ in the large-system limit, 
and that empirical eigenvalue distribution of $R$ 
converges to a limiting eigenvalue distribution $\rho(\lambda)$ with a finite support 
included in $[\lambda_{\min},\,\lambda_{\max}]$, in the large-system limit.
Under the assumption of random spreading,
we consider the average conditional mutual information in the large-system limit,
\begin{equation}
C=\lim_{K\to\infty}\mathbb{E}_S\{C_{\mathrm{user}}\}
=\lim_{K \rightarrow \infty} \frac{1}{K}I(\bm{x};\bm{y}|S), \label{eq:capacity}
\end{equation}
where $\mathbb{E}_S$ denotes expectation with respect to $S$, and 
where $I(\bm{x};\bm{y}|S)$ is conditional mutual information between $\bm{x}$ and $\bm{y}$ given $S$.  
We discuss maximization of $C$ 
with respect to characteristics of the random matrix $S$.
\section{Analysis}
\subsection{Evaluation of average mutual information via replica method}
The average mutual information~\eqref{eq:capacity} is decomposed 
into two terms, 
\begin{align}
C&=\lim_{K\rightarrow \infty } \frac{1}{K} \left[\mathbb{E}_{\bm{y},S}\{\log p(\bm{y}|S)\}-\mathbb{E}_{\bm{y},\bm{x},S}\{ \log p(\bm{y}|\bm{x},S)\}\right]\\
 &=\mathcal{F} - \frac{1}{2 \beta} \left( 1+ \log (2\pi \sigma^2)\right), \label{eq:capa2}
\end{align}
with 
\begin{equation}
\mathcal{F}\equiv -\lim_{K \rightarrow \infty} \frac{1}{K} \mathbb{E}_{\bm{y}, S}\{\log p(\bm{y}|S)\}, \label{eq:free-energy}
\end{equation}
where $\mathbb{E}_{\bm{y},S}$ denotes expectation with respect to $\bm{y}$ and $S$.
Direct calculation of the right-hand side of~\eqref{eq:free-energy} is 
in general computationally intractable.
In order to evaluate~\eqref{eq:free-energy}, we invoke the replica method.
Substituting the identity 
\begin{equation}
\lim_{n \rightarrow 0} \frac{\partial}{\partial n} (p(\bm{y}|S))^n = \log p(\bm{y} |S)
\end{equation}
to the right-hand side of~\eqref{eq:free-energy}, we obtain 
\begin{equation}
\mathcal{F}
 = -\lim_{K \rightarrow \infty} \frac{1}{K} 
 \lim_{n \rightarrow 0}\frac{\partial}{\partial n} \log {\mathbb E}_{\bm{y},S}\{ (p(\bm{y}|S))^n\}. \label{eq:F_1}
 \end{equation}
We assume that the limit with respect to $K$ 
and the limit and the differentiation with respect to $n$ 
are interchangeable without altering the final result, obtaining 
 \begin{equation}
\mathcal{F} = -\lim_{n \rightarrow 0} \frac{\partial}{\partial n}
\lim_{K \rightarrow \infty}\frac{1}{K} \log {\mathbb E}_{\bm{y},S}\{ (p(\bm{y}|S))^n\}.
\label{eq:post_replica}
\end{equation}
The limit $K\to\infty$ allows us to apply the saddle-point method 
to evaluate a relevant quantity.  
We apply replica trick in order to evaluate~\eqref{eq:post_replica}, 
in which we first evaluate ${\mathbb E}_{\bm{y},S}\{(p(\bm{y}|S))^n \}$ 
assuming that $n$ is a non-negative integer, 
and then perform the limit and the differentiation with respect to $n$, 
assuming that $n$ is real. 

Evaluation of~\eqref{eq:post_replica} basically goes in a similar manner as~\cite{tuk}.
Detailed analysis is described in the appendix.
Here, we only show the result.
The average mutual information in the large-system limit is given by
\begin{equation}
C = -\frac{1}{2}\theta\mathcal{E} - \frac{1}{2}G\left(-\frac{\mathcal{E}}{\sigma^2}\right) 
- \frac{1}{2} \log \frac{2 \pi}{\theta}
-\frac{1}{2} - \int p(u; \theta)\log p(u; \theta)\,\mathrm{d}u,
\label{eq:result}
\end{equation}
where $\{\mathcal{E},\,\theta\}$ are parameters 
whose values are to be determined later, and where $p(u; \theta)$ 
is a probability density function of output $u$ 
of a scalar AWGN channel with $1/\theta$ the noise variance, 
when the channel input $x$ is generated from the distribution $p(x)$. 
The function $G(t)$ is defined as 
\begin{equation}
G(t)=\int_0^t {\bf R}(z)\,\mathrm{d}z \label{eq:G}
\end{equation}
where ${\bf R}(z)$ is the R-transform~\cite{GM} 
of the limiting eigenvalue distribution $\rho(\lambda)$ 
of the correlation matrix $R$, which is defined 
on the basis of the Hilbert transform\footnote{%
It should be noted that the so-called Cauchy transform is defined 
by the same formula as the Hilbert transform~\protect\eqref{eq:Hilbert-trans}, 
but with $\gamma$ in the upper half of complex plane.}
of $\rho(\lambda)$, 
\begin{equation}
\mathcal{C}(\gamma) = \int \frac{\rho(\lambda)}{\gamma -\lambda}\,\mathrm{d}\lambda,\quad \gamma<\lambda_{\min},
\label{eq:Hilbert-trans}
\end{equation}
as 
\begin{equation}
\mathcal{C}\left({\bf R}(z)+ \frac{1}{z}\right)= z.
\label{eq:R-trans}
\end{equation}
The parameters $\{\mathcal{E},\,\theta \}$ are to be determined 
from the following saddle-point equations:
\begin{align}
\mathcal{E}&={\mathbb E} \{ (x-\langle x \rangle)^2 ;\,\theta \}
\label{eq:fix1}
\\
\theta &= \frac{1}{\sigma^2} {\bf R}\left(- \frac{\mathcal{E}}{\sigma^2}\right)
\label{eq:fix2}
\end{align}
where $\langle x\rangle$ denotes posterior mean estimate 
of the channel input $x$ of the scalar AWGN channel introduced above, 
defined as 
\begin{equation}
\langle x\rangle
=\frac{{\displaystyle\int}x\sqrt{\frac{\theta}{2\pi}}e^{-\theta(u-x)^2/2}\,p(x)\,\mathrm{d}x}%
{{\displaystyle\int}\sqrt{\frac{\theta}{2\pi}}e^{-\theta(u-x)^2/2}\,p(x)\,\mathrm{d}x},
\end{equation}
and where $\mathbb{E}$ in~\eqref{eq:fix1} denotes expectation 
with respect to the channel input $x$ and output $u$ 
of the scalar AWGN channel.  

\begin{figure}[htbp]
\includegraphics[width=0.45\textwidth]{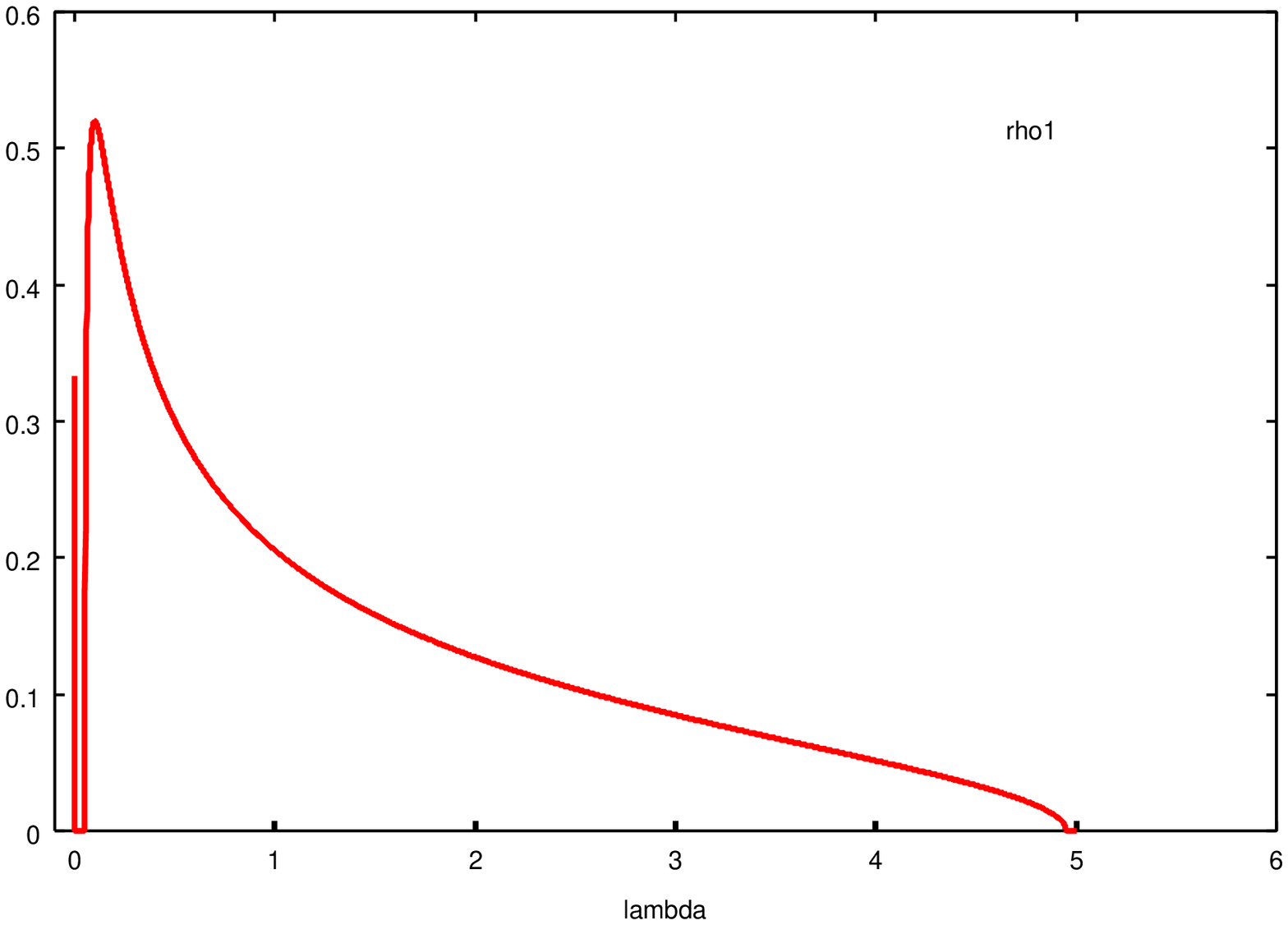}
\hspace{0.05\textwidth}
\includegraphics[width=0.45\textwidth]{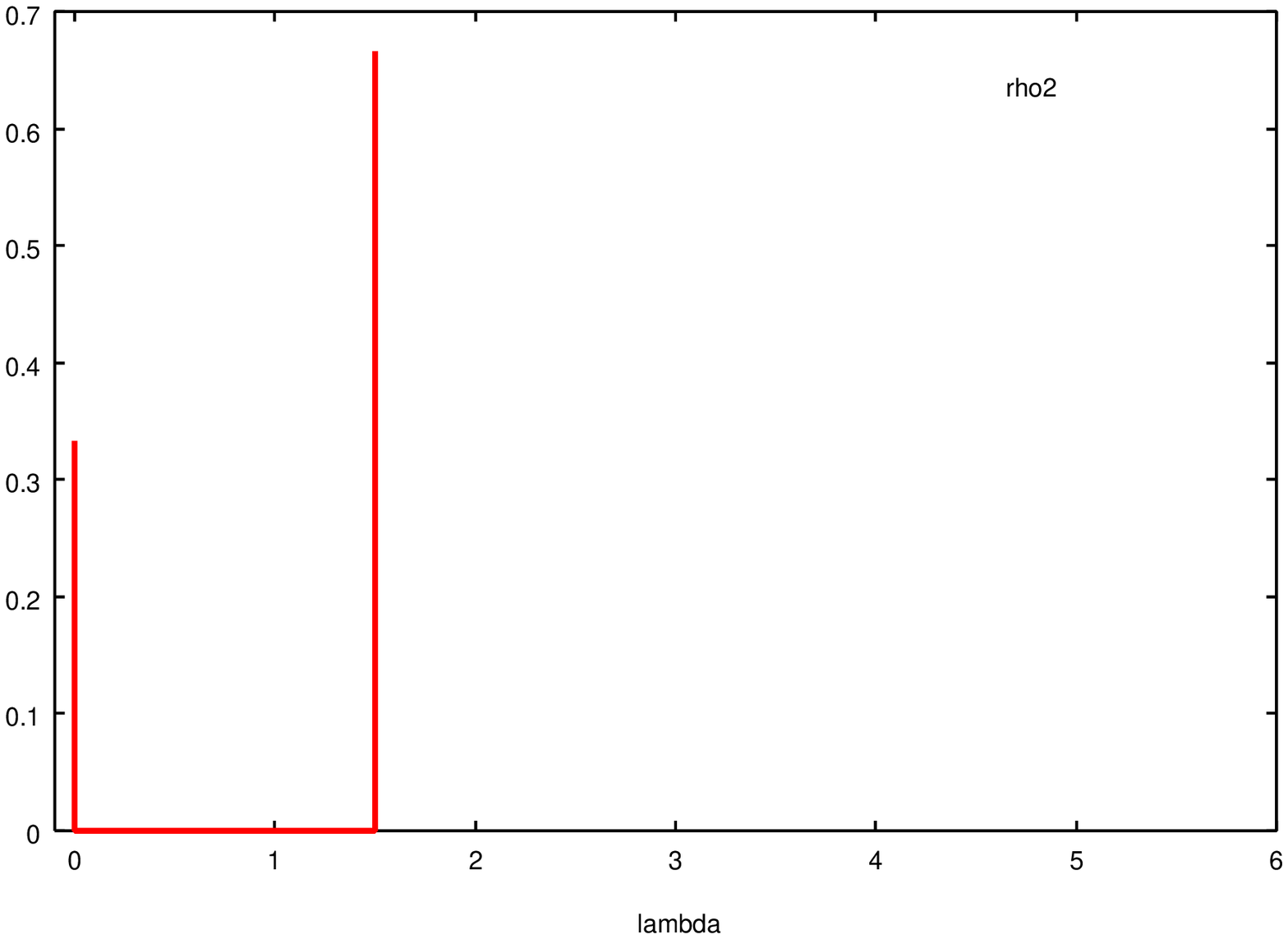}
\caption{Eigenvalue distributions $\rho^{\mathrm{MP}}$(left figure) and 
$\rho^{\mathrm{WBE}}$(right figure) with $\beta=1.5$. Both distributions have 
probability weight $1-1/\beta$ at $\lambda=0$.}
\label{fig:eig}
\end{figure}
\begin{figure}[htbp]
\begin{center}
\includegraphics[height=0.3\textheight]{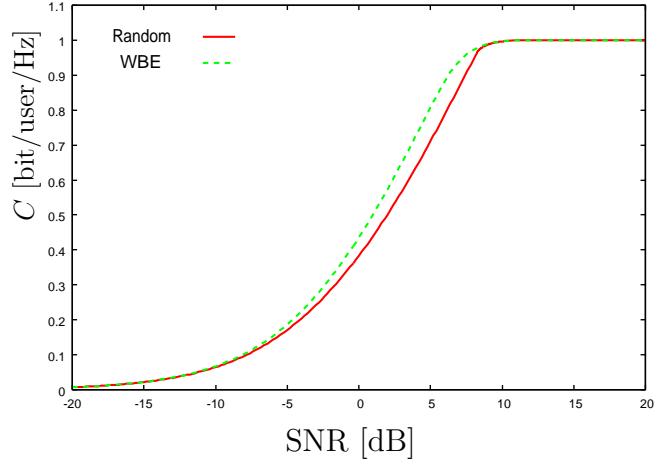}
\caption{The average mutual information $C$ when $\rho^{\mathrm{WBE}}$ (WBE) and $\rho^{\mathrm{MP}}$ (Random)
are specified, in case of $\beta=1.5$.}
\label{fig:Random-WBE}
\end{center}
\end{figure}

The average mutual information~\eqref{eq:result} depends on the limiting eigenvalue distribution $\rho(\lambda)$ of $R$ 
as well as the prior distribution $p(x_k)$.
When we assume that $s_{\mu k}$ are i.i.d.\ random variables 
whose mean and variance are zero and one, respectively, 
our result is reduced to that obtained by Guo and Verd\'u~\cite{guo}.
In this case, the limiting eigenvalue distribution 
$\rho(\lambda)$ is given by the so-called Mar\u{c}enko-Pastur law~\cite{random},
\begin{equation}
\rho^{\mathrm{MP}}(\lambda) = \left( 1-\frac{1}{\beta}\right)^{+}\delta(\lambda) + 
\frac{\sqrt{(\lambda - a)^+(b-\lambda)^+}}{2\pi \beta \lambda}, \label{eq:Mar}
\end{equation}
where $(x)^+=\max (0,x)$, and $a=(1-\sqrt{\beta})^2$, $b=(1+\sqrt{\beta})^2$ 
(see figure~\ref{fig:eig}), whose 
R-transform is given by
\begin{equation}
{\bf R}^{\mathrm{MP}}(z)= \frac{1}{1-\beta z}. \label{eq:Mar-R}
\end{equation}
Substituting~\eqref{eq:Mar-R} to~\eqref{eq:G}, one can confirm the above-mentioned fact.
Our analysis also includes the case of the system with WBE spreading sequences.
Since the characteristic of WBE spreading sequences is expressed as~\eqref{eq:WBE},
the correlation matrix of WBE spreading sequences has trivial zero eigenvalue and 
the eigenvalue $\lambda=\beta$, with multiplicities $(K-L)$ and $L$, respectively 
(see figure~\ref{fig:eig}), and therefore 
\begin{equation}
\rho^{\mathrm{WBE}}(\lambda) = \left( 1- \frac{1}{\beta}\right) \delta (\lambda) + \frac{1}{\beta} \delta (\lambda -\beta), \label{eq:WBE-eig}
\end{equation}
whose R-transform is given by
\begin{equation}
{\bf R}^{\rm WBE}(z)=\frac{2}{1-\beta z + \sqrt{(\beta z -1)^2 +4z}}. \label{eq:R-wbe}
\end{equation}
One can evaluate, via ${\bf R}^{\rm WBE} (z)$, the mutual information 
when orthogonally-invariant random WBE spreading sequences are 
employed.
In figure~\ref{fig:Random-WBE}, we show a comparison 
of the mutual information when the above two spreading sequences
are employed, and when probability distribution of $\{ x_k \}$ is given by 
$p(x_k)=(\delta(x_k-1)+\delta(x_k+1))/2$, $k=1,\cdots,K$.
One can confirm that WBE spreading sequences achieve higher mutual information 
than i.i.d.\ random spreading sequences do.
\subsection{Optimizing spreading sequences}
Choices of spreading sequences affect the average mutual information $C$ 
through the limiting eigenvalue distribution $\rho(\lambda)$ of the correlation matrix $R$.
Then, we regard the average mutual information $C$ as a functional of $\rho(\lambda)$,
and seek the eigenvalue distribution $\rho^*(\lambda)$
which maximizes the average mutual information $C$.
Hereafter, we consider the case of $\beta>1$ since optimal spreading sequences in the case of $\beta\le1$
are obviously orthogonal spreading sequences.
In optimizing $C$ with respect to $\rho(\lambda)$, 
the following two constraints should be imposed on $\rho(\lambda)$: 
First, since $\beta>1$, the $K\times K$ matrix $R$ has trivial zero eigenvalues with multiplicity $(K-L)$. 
Second, since we have normalized the power of spreading sequences as $\sum_{\mu=1}^L (s_{\mu k}/\sqrt{L})^2 =1 , k=1,\cdots, K$, 
the matrix $S$ should satisfy
\begin{equation}
\mathrm{Tr} S^TS =\sum_{k=1}^K \lambda_k= K, \label{eq:cond-normsp}
\end{equation}
where $\{\lambda_k\}$ are the eigenvalues of $K \times K$ matrix $S^TS$.
In terms of $\rho(\lambda)$, the constraint~\eqref{eq:cond-normsp} is expressed as
\begin{equation}
\int \lambda\,\rho (\lambda)\,\mathrm{d}\lambda = 1. \label{eq:cond-ave-rho}
\end{equation} 
We rewrite $\rho(\lambda)$ in view of these constraints as 
\begin{equation}
\rho (\lambda) = \left(1-\frac{1}{\beta}\right) \delta(\lambda) + \frac{1}{\beta} \pi (\lambda) \label{eq:rho-pi}, 
\end{equation}
where $\pi (\lambda)$ satisfies  
\begin{equation}
\int \pi (\lambda)\,\mathrm{d}\lambda =1, \label{eq:norm_prob}
\end{equation}
as the normalization as a probability distribution, and
\begin{equation}
\int \lambda\,\pi (\lambda)\,\mathrm{d}\lambda = \beta, \label{eq:norm_power}
\end{equation}
which corresponds to the normalization of the power of spreading sequences~\eqref{eq:cond-ave-rho}.

In order to discuss the extremum of $C$ with respect to $\rho (\lambda)$,
 we consider first-order perturbations of $C$.
Since the parameters affected by the perturbation of $\rho (\lambda)$ are 
$\{G(t),\, \mathcal{E},\, \theta\}$, 
the functional derivative of $C$ with respect to $\rho$ is expressed as 
\begin{equation}
\frac{\delta C}{\delta \rho} = \frac{\delta C}{\delta G}\cdot\frac{\delta G}{\delta \rho} 
+ \frac{\partial C}{\partial\mathcal{E}}\cdot\frac{\delta \mathcal{E}}{\delta \rho}
+ \frac{\partial C}{\partial\theta}\cdot\frac{\delta \theta}{\delta \rho}. 
\end{equation}
Since $\{\mathcal{E},\,\theta\}$ should satisfy 
the saddle-point equations~\eqref{eq:fix1} and \eqref{eq:fix2},
the derivatives of $C$ with respect to the parameters $\{\mathcal{E},\,\theta\}$ should be zero
at the saddle point.  
Therefore, one can safely ignore the effects of perturbations via 
$\mathcal{E}$ and $\theta$.  

Our next observation is that, 
if one can find an eigenvalue distribution which maximizes $-(1/2)G(-\mathcal{E}/\sigma^2)$,
 which is the only term having the first-order effect in~\eqref{eq:result}, 
it also maximizes the average mutual information~\eqref{eq:result}.
We rewrite $-G(-\mathcal{E}/\sigma^2)$ as
\begin{equation}
-G\left(-\frac{\mathcal{E}}{\sigma^2}\right)=-\int_0^{-\mathcal{E}/\sigma^2} {\bf R}(z) 
\mathrm{d}z=\int_{-\mathcal{E}/\sigma^2}^0 {\bf R}(z)\mathrm{d}z.
\end{equation}
Since $-\mathcal{E}/\sigma^2 <0$ and ${\bf R}(z) > 0$, one can make the following statement:
If there is a distribution $\rho^* (\lambda)$ whose R-transform ${\bf R}^* (z)$ satisfies
\begin{equation}
{\bf R}^* (z) \ge {\bf R} (z), \quad\text{for}\,\,{}^{\forall}z\in (-\mathcal{E}/\sigma^2, 0), \label{eq:opt-cond}
\end{equation}
for any R-transform ${\bf R}(z)$ of the distribution $\rho(\lambda)$ 
which satisfies the constraints~\eqref{eq:rho-pi}--\eqref{eq:norm_power},
$\rho^* (\lambda)$ also maximizes $-G(-\mathcal{E}/\sigma^2)$.
We summarize the above arguments in the next proposition. 
\begin{proposition} If one can find an eigenvalue distribution $\rho^*(\lambda)$ which 
maximizes R-transform for ${}^{\forall}z\in(-\mathcal{E}/\sigma^2, 0)$, 
$\rho^*(\lambda)$ also maximizes the average mutual information $C$.
\end{proposition}
It should be noted that the existence of $\rho^* (\lambda)$ is not guaranteed at this stage.
However, in the following, we show that there is a distribution 
which satisfies the condition of Proposition 1.  

As a next step, we convert the optimization problem in terms of R-transform 
into the one in terms of Hilbert transform.
Since Hilbert transform $\mathcal{C}(\gamma)$ is a monotonically decreasing function of $\gamma$, and 
since Hilbert transform has the relation~\eqref{eq:R-trans} with R-transform, 
it follows that ${\bf R}(z) + 1/z$ also decreases monotonically with respect to $z$.
This fact leads us to the following statement: If one can find an eigenvalue distribution
$\rho^*(\lambda)$ which maximizes $\mathcal{C} (\gamma)$ for ${}^{\forall}\gamma<\lambda_{\mathrm{min}}$,
 and which satisfies~\eqref{eq:rho-pi}--\eqref{eq:norm_power},
  $\rho^*(\lambda)$ also maximizes ${\bf R}(z)$ for ${}^{\forall}z\in (z_{\mathrm{min}},\,0)$, with 
\begin{equation}
z_{\mathrm{min}}= \lim_{\gamma \rightarrow \lambda_{\mathrm{min}}-0} \mathcal{C}(\gamma).
\end{equation}
Summarizing the arguments so far, we can state the next proposition.
\begin{proposition}
If one can find an eigenvalue distribution $\rho^*(\lambda)$ 
whose Hilbert transform $\mathcal{C}^*(\gamma)$ satisfies
\begin{equation}
\mathcal{C}^*(\gamma) \ge \mathcal{C}(\gamma),\quad {}^{\forall}\gamma < \lambda_{\min},
\end{equation}
for any eigenvalue distribution $\rho(\lambda)$ 
with Hilbert transform $\mathcal{C}(\gamma )$, 
which satisfies~\eqref{eq:rho-pi}--\eqref{eq:norm_power}, $\rho^*(\lambda)$ also maximizes the 
average mutual information $C$.
\end{proposition}

Following the above proposition, we consider the maximization problem 
of the Hilbert transform for $\gamma < \lambda_{\min}$.  
Substituting~\eqref{eq:rho-pi} to~\eqref{eq:Hilbert-trans}, 
the Hilbert transform $\mathcal{C}(\gamma)$ is rewritten as 
\begin{equation}
\mathcal{C}(\gamma) = \left(1- \frac{1}{\beta}\right) \frac{1}{\gamma}
+ \frac{1}{\beta} \int \frac{\pi (\lambda)}{\gamma-\lambda}\mathrm{d}\lambda.
\label{eq:Hilbert-2}
\end{equation}
Since the first term, which is derived from the trivial zero eigenvalues, 
has no room for optimization, 
we maximize the second term under the constraints~\eqref{eq:norm_prob} 
and \eqref{eq:norm_power}.
Let us consider the following integral: 
\begin{equation}
\int \pi(\lambda)\left(\frac{1}{\gamma - \lambda} - f(\lambda)\right)\mathrm{d}\lambda, 
\label{eq:modified-obj}
\end{equation}
where $f(\lambda)$ is a linear function 
tangential to $1/(\gamma - \lambda)$ at $\lambda = \beta$. 
Since the function $f(\lambda)$ is a linear function, 
the expectation of $f(\lambda)$ with respect to $\pi(\lambda)$ is constant 
under the constraints~\eqref{eq:norm_prob} and \eqref{eq:norm_power}.
Thus, the quantities~\eqref{eq:modified-obj} and~\eqref{eq:Hilbert-2} 
are maximized by the same eigenvalue distribution.
We here consider a maximization problem of the objective function~\eqref{eq:modified-obj} for $\gamma < \lambda_{\min}$
with the constraint~\eqref{eq:norm_prob}, 
but {\it without} the constraint~\eqref{eq:norm_power}. 
Since we have only to consider $\lambda \in [\lambda_{\min}, \lambda_{\max}]$, 
we can assume $\lambda > \gamma$. 
Since $1/(\gamma -\lambda)$ is convex upward in $\lambda$ for $\lambda > \gamma$,
one has 
\begin{equation}
\frac{1}{\gamma - \lambda} -f(\lambda) \le 0,\quad \lambda > \gamma,
\end{equation}
where the equality holds if and only if $\lambda =\beta$.
Then, the objective function~\eqref{eq:modified-obj} is maximized for $\gamma < \lambda_{\min}$ by the probability distribution
\begin{equation}
\pi(\lambda)=\delta (\lambda-\beta). \label{eq:opt}
\end{equation}
Since the distribution~\eqref{eq:opt} incidentally satisfies the power constraint~\eqref{eq:norm_power}, 
the distribution~\eqref{eq:opt} is also a maximizer of the objective function~\eqref{eq:modified-obj} 
with both of the constraints~\eqref{eq:norm_prob} and~\eqref{eq:norm_power}. 
Since the two functions~\eqref{eq:Hilbert-2} and~\eqref{eq:modified-obj} 
are maximized by the same probability distribution, 
the distribution~\eqref{eq:opt} is also the optimal solution of the maximization problem 
of the Hilbert transform. Thus, we obtain the maximizer of the average mutual 
information~\eqref{eq:result}.
Substituting~\eqref{eq:opt} to~\eqref{eq:rho-pi}, 
one can confirm that the optimal eigenvalue distribution is 
the one of WBE spreading sequences $\rho^{\mathrm{WBE}}$, 
which is given by~\eqref{eq:WBE-eig}.

We have so far shown that WBE spreading sequences are 
also asymptotically optimal in CDMA systems with a non-Gaussian input distribution 
in the large-system limit.  
This finding is an extension of the optimality result 
of WBE spreading sequences for Gaussian-input CDMA systems.  
\section{Conclusion}
We have demonstrated that the statistical-mechanics approach 
is useful not only in analyzing theoretical performance 
of wireless communication systems 
but also in providing clues to how to design them, 
via the problem of optimizing spreading sequences in CDMA systems. 
We have evaluated, via replica method, average mutual information between input and 
output of the system in the large-system limit, and discussed 
the optimization problem of the average mutual information 
in terms of characteristics of random spreading sequences.  
It has been shown that the average mutual information is maximized 
in the large-system limit by orthogonally-invariant random WBE spreading sequences 
even when the inputs are non-Gaussian.  
Although in this paper we have only studied a fully-synchronous CDMA model with perfect power control, 
one can consider the same problem in more general CDMA systems, such as the one with 
unequal-power users, and that is deferred to our future work.
\appendix
\section{Details of replica analysis}
\label{appA}
In this appendix, we explain how to evaluate $\mathcal{F}$ given by~\eqref{eq:post_replica}.
First, we calculate the expectation 
$\mathbb{E}_{\bm{y},S} \{p(\bm{y}|S)^n\}$ assuming that $n$ is 
a non-negative integer.
Introducing replicated random vectors $\bm{x}_a=[x_{a1},\cdots, x_{aK}]^T \in 
\mathbb{R}^K,\,a=0,\cdots, n$,
 which are drawn from the same probability distribution as $\bm{x}$,
we rewrite $\mathbb{E}_{\bm{y},S} \{p(\bm{y}|S)^n\}$ as
\begin{equation}
\mathbb{E}_S
\left\{\int \int \prod_{a=0}^n p(\bm{y}|\bm{x}_a,S)p(\bm{x}_a )\mathrm{d}\bm{x}_a \mathrm{d}\bm{y}\right\}.
\end{equation}
Performing the integral with respect to $\bm{y}$, we obtain
\begin{multline}
\mathbb{E}_{\bm{y},S}\{(p(\bm{y}|S))^n\}\\ 
=\mathbb{E}_{\bm{x}_a,S}\bigg\{\exp\left[\frac{K}{2}\mathrm{Tr}RV - \frac{L}{2}\log
(n+1) - \frac{nL}{2}\log(2\pi \sigma^2)\right]\bigg\}.
\end{multline}
where $K\times K$ matrix $V$ is given by
\begin{equation}
V=\frac{1}{(n+1)K\sigma^2}\left(\sum_{a=0}^n\bm{x}_a\right)\left(\sum_{a=0}^n\bm{x}_a\right)^T
-\frac{1}{K\sigma^2}\sum_{a=0}^n \bm{x}_a\bm{x}_a^T.
\end{equation}
The expectation with respect to $S$ can be performed 
via the so-called Itzykson-Zuber integral~\cite{iz,GM} (see also~\cite{tuk,tanaka2008}), 
since $R=S^TS$ is assumed orthogonally invariant 
and rank of $V$ is at most $(n+1)$, as 
\begin{equation}
\lim_{K\to\infty}\frac{1}{K}\log\mathbb{E}_S\left\{\exp\left[\frac{K}{2}\mathrm{Tr}RV\right]\right\}
=\frac{1}{2}\mathrm{Tr}G(V),
\end{equation}
where $G(x)$ is the function defined in~\eqref{eq:G}. 
Thus, we obtain the following equation, ignoring vanishing terms in the large-system limit, 
\begin{multline}
\mathbb{E}_{\bm{y},S}\{(p(\bm{y}|S))^n\} \\
 = \mathbb{E}_{\{\bm{x}_a\}}\left\{\exp\left[\frac{K}{2}\mathrm{Tr}G(V)
 - \frac{L}{2}\log
(n+1) - \frac{nL}{2}\log(2\pi \sigma^2)\right]\right\}. \label{eq:moment}
\end{multline}

We next take expectation of~\eqref{eq:moment} with respect to $\{\bm{x}_a\}$.  
Since eigenvalues of the matrix $V$ are functions of $\{\bm{x}_a\}$ 
only through their inner products $\bm{x}_a \cdot \bm{x}_b,\,a, b=0,\cdots, n$, 
we rewrite the expectation with respect to $\{\bm{x}_a\}$ into the one 
with respect to the $(n+1)\times (n+1)$ matrix
\begin{equation}
\label{eq:em}
Q=(Q_{ab}),\quad Q_{ab}= \frac{1}{K} \sum_{k=1}^K  x_{ak}x_{bk},
\end{equation}
as 
\begin{equation}
\int \exp [K\mathcal{G}(Q)]\mu_K(Q)\mathrm{d}Q, \label{eq:pre-saddle}
\end{equation}
where $K\mathcal{G}(Q)$ is the exponent of~\eqref{eq:moment},
\begin{equation}
\mathcal{G}(Q)=
\left[\frac{1}{2}\mathrm{Tr}G(V)\right](Q)-\frac{1}{2\beta}\log(n+1)-\frac{n}{2\beta}\log(2\pi\sigma^2),
\end{equation}
and where $\mu_K$ is the following measure
\begin{equation}
\mu_K(Q)=\mathbb{E}_{\{\bm{x}_a\}}\left\{\prod_{0\le a \le b}^n\delta \left(\sum_{k=1}^K x_{ak}x_{bk}-KQ_{ab}\right)\right\}. 
\end{equation}
Utilizing the saddle-point method~\cite{tt,guo}, 
we evaluate~\eqref{eq:pre-saddle} in the limit $K\rightarrow \infty$ as
\begin{equation}
\lim_{K\rightarrow \infty}\frac{1}{K}\log \mathbb{E}\{(p(\bm{y}|S))^n\}
=\sup_{Q}\{\mathcal{G}(Q)-\mathcal{I}(Q)\}, \label{eq:saddle-point-method}
\end{equation}
where $\mathcal{I}(Q)$ is the rate function of the empirical means~\eqref{eq:em}, defined via 
a Legendre transform as 
\begin{equation}
\mathcal{I}(Q)=\sup_{\tilde{Q}}\left[\sum_{0\le a \le b} Q_{ab}\tilde{Q}_{ab}-\log M(\tilde{Q})\right],
\end{equation}
where $\tilde{Q}=(\tilde{Q}_{ab})$ is an $(n+1)\times(n+1)$ symmetric matrix.
The cumulant generating function $\log M(\tilde{Q})$ of $\{x_a\}$ is defined as
\begin{equation}
\log M(\tilde{Q})=\log \mathbb{E}_{\{x_a\}}\left\{\exp\left[\sum_{0\le a\le b}
\tilde{Q}_{ab} x_{a}x_{b}\right]\right\}.
\end{equation}

In order to proceed further, we assume the so-called {\it replica symmetry}: 
 We assume that the extremum of~\eqref{eq:saddle-point-method} 
is invariant under exchanges of the replica indexes. 
Under the assumption of replica symmetry, we introduce new parameters, 
\begin{equation}
Q_{aa}=p,\quad Q_{ab}=q,\quad a\not=b.
\end{equation}
Using these parameters, the eigenvalues of $V$ are expressed as
\begin{equation}
\lambda_1= -\frac{p-q}{\sigma^2}, \label{eq:l1}
\end{equation}
\begin{equation}
\lambda_2=0,
\end{equation}
whose multiplicities are $n$ and $(K-n)$, respectively.
Since $G(x)$ is an analytic function and $G(0)=0$, we can express $\mathcal{G}(Q)$ as 
\begin{equation}
\mathcal{G}(Q)=\frac{n}{2}G\left(-\frac{p-q}{\sigma^2}\right)
-\frac{1}{2\beta}\log(n+1)-\frac{n}{2\beta}\log(2\pi\sigma^2). \label{eq:func-gq}
\end{equation}
We also apply the assumption of replica symmetry to $\tilde{Q}$ as 
\begin{equation}
\tilde{Q}_{aa}=c,\quad \tilde{Q}_{ab}=\theta,\quad a\not=b,
\label{eq:til-Q}
\end{equation}
and rewrite the rate function $\mathcal{I}(Q)$ as 
\begin{multline}
\mathcal{I}(Q)= 
\sup_{\tilde{Q}}\bigg[ (n+1)cp + \frac{n(n+1)}{2}\theta q\\
-\log \mathbb{E}_{\{x_a\}} \bigg\{\exp\left[c\sum_{a=0}^n x_a
+ \theta \sum_{0\le a < b}x_a x_b\right]\bigg\} \bigg]. \label{eq:rate-result}
\end{multline}
Deriving the extremum condition of~\eqref{eq:saddle-point-method} 
with respect to $p$ and $q$ in the limit $n\rightarrow 0$, we obtain the following saddle-point equations,
\begin{equation}
c=0, \label{eq:cond-c}
\end{equation}
\begin{equation}
\theta =\frac{1}{\sigma^2}{\bf R}(-\frac{p-q}{\sigma^2}). \label{eq:cond-theta}
\end{equation}
Similarly, as the extremum condition of~\eqref{eq:saddle-point-method}, 
we obtain the following saddle-point equation,
\begin{equation}
p-q=\mathbb{E}\{(x-\langle x \rangle)^2; \theta\}, \label{cond-pq}
\end{equation}
where $x$ follows the probability distribution $p(x)$, 
and where $\langle x \rangle$ is the posterior mean estimate of 
channel input $x$ given output $u$ 
in a single-user AWGN channel whose variance is $1/\theta$ and whose input and output are $x$ and $u$, 
respectively.
Introducing the posterior variance $\mathcal{E}\equiv p-q$, 
we obtain the saddle-point equations~\eqref{eq:fix1} and \eqref{eq:fix2}.
In order to obtain $\mathcal{F}$, we differentiate 
$\log \mathbb{E}\{(p(\bm{y}|S))^n\}$, which is obtained by 
substituting~\eqref{eq:func-gq} and~\eqref{eq:rate-result} 
to~\eqref{eq:saddle-point-method}, with respect to $n$, 
and then take the limit $n \rightarrow 0$. 
Finally, we obtain the representation of $\mathcal{F}$ as
\begin{multline}
\mathcal{F}=
 -\frac{1}{2}\theta\mathcal{E} - \frac{1}{2}G\left(-\frac{\mathcal{E}}{\sigma^2}\right)
  - \frac{1}{2} \log \frac{2 \pi}{\theta}\\
 -\frac{1}{2} - \int p(u; \theta)\log p(u; \theta)\,\mathrm{d}u + \frac{1}{2\beta}(1+\log(2\pi \sigma^2)). 
 \label{eq:result-F}
\end{multline}
Substituting~\eqref{eq:result-F} to~\eqref{eq:capa2}, we obtain the average mutual information~\eqref{eq:result}.
\section*{Acknowledgment}
Support from the Grant-in-Aid for Scientific Research on Priority Areas, 
the Ministry of Education, Culture, Sports, Science and Technology, Japan 
(no.~18079010) is acknowledged.

\end{document}